\documentstyle[11pt,epsfig]{article}
\title{\bf Accelarating universe in brane gravity with confining potential}
\author{M. Heydari-fard, M. Shirazi,  S. Jalalzadeh$^1$\thanks{email:
s-jalalzadeh@sbu.ac.ir}  and H. R. Sepangi$^{1,2}$\thanks{email:
hr-sepangi@sbu.ac.ir}
\\ $^1${\small Department of Physics, Shahid Beheshti University, Evin, Tehran 19839, Iran}\\$^2${\small
Institute for Studies in Theoretical Physics and Mathematics, P.O.
Box 19395-5746, Tehran, Iran }}
\begin{document}
\maketitle 
\begin{abstract}
We construct the Einstein field equations on a 4-dimensional brane
embedded in an $m$-dimensional bulk where the matter fields are
confined to the brane by means of a confining potential. As a
result, an extra term in the Friedmann equation in an
$m$-dimensional bulk appears that may be interpreted as the
X-matter, providing a possible phenomenological explanation for
the acceleration of the universe. The study of the relevant
observational data suggests good agreement with the predictions of
this model.
\vspace{5mm}
\end{abstract}
\section{Introduction}
There has been considerable activity in the recent past in the
area of higher dimensional gravity where the classical $4D$
space-time of General Relativity is recovered as an effective
theory \cite{1}. The basic idea in these theories is the existence
of a higher dimensional bulk in which our universe, called the
brane, is sitting as a hypersurface. Physical matter fields are
confined to this hypersurface, while gravity can reside in the
higher dimensional space-time as well as on the brane. This
paradigm was first proposed as a means to reconcile the mismatch
between the scales of particle physics and gravity \cite{2}. The
cosmological evolution of such a brane universe has been
extensively investigated. Exact solutions have been found by
several authors \cite{3,4,5}. These solutions reduce to a
generalized Friedmann equation on the brane which contains a
quadratic term related to matter energy density. This term arises
from the imposition of the Israel junction conditions which is a
relationship between the extrinsic curvature and energy-momentum
tensor of the brane and results from the singular behavior in the
energy-momentum tensor. The main difficulty in applying a junction
condition is that it is not unique. Other forms of junction
conditions exist, so that the different conditions may lead to
different physical results \cite{6}. Furthermore, these conditions
cannot be used when more than one non-compact extra dimension is
involved. An interesting higher-dimensional model was introduced
in \cite{7} where particles are trapped in a 4-dimensional
hypersurface by the action of the confining potential $\cal V$.
Also, in \cite{8}, the dynamics of test particles confined to a
brane by the action of a confining potential, at classical and
quantum levels have been studied and the effects of small
perturbations along the extra dimensions investigated. Within the
classical limits, the particle remains stable under small
perturbations and the effects of the extra dimensions are not felt
by the test particle, hence making them undetectable in this way.
The quantum fluctuations of the brane cause the mass of a test
particle to become quantized and, interestingly, the Yang-Mills
fields appear as quantum effects.

In this paper, we follow \cite{8} and consider an $m$-dimensional
bulk space without imposing the $Z_2$ symmetry. To localize the
matter on the brane, a confining potential is used rather than a
delta-function in the energy-momentum tensor. As a result, the
extrinsic curvature becomes independent of the matter content of
the brane. The Friedmann equation in this model is modified by the
appearance of an extra term which behaves like the X-matter; the
phenomenological model proposed to fit the data explaining
accelerated expansion of the universe. We should emphasize here
that there is a difference between the model presented in this
work and models introduced in \cite{9,10} in that in the latter no
mechanism is introduced to account for the confinement of matter
to the brane.
\section{The model }
In this section we present a brief review of the model proposed in
\cite{8}.  Consider the background manifold $ \overline{V}_{4} $
isometrically embedded in a pseudo-Riemannian manifold $ V_{m}$ by
the map ${ \cal Y} : \overline{V}_{4}\rightarrow  V_{m} $ such
that
\begin{eqnarray}\label{a}
{\cal G} _{AB} {\cal Y}^{A}_{,\mu } {\cal Y}^{B}_{,\nu}=
\bar{g}_{\mu \nu}  , \hspace{.5 cm} {\cal G}_{AB}{\cal
Y}^{A}_{,\mu}{\cal N}^{B}_{a} = 0  ,\hspace{.5 cm}  {\cal
G}_{AB}{\cal N}^{A}_{a}{\cal N}^{B}_{b} = g_{ab}= \pm 1.
\end{eqnarray}
where $ {\cal G}_{AB} $  $ ( \bar{g}_{\mu\nu} ) $ is the metric of
the bulk (brane) space  $  V_{m}  (\overline{V}_{4}) $ in
arbitrary coordinates, $ \{ {\cal Y}^{A} \} $   $  (\{ x^{\mu} \})
$  is the  basis of the bulk (brane) and  ${\cal N}^{A}_{a}$ are
$(m-4)$ normal unite vectors, orthogonal to the brane. The
perturbation of $\bar{V}_{4}$ in a sufficiently small neighborhood
of the brane along an arbitrary transverse direction $\xi$ is
given by
\begin{eqnarray}\label{a1}
{\cal Z}^{A}(x^{\mu},\xi^{a}) = {\cal Y}^{A} + ({\cal
L}_{\xi}{\cal Y})^{A}, \label{eq2}
\end{eqnarray}
where $\cal L$ represents the Lie derivative. By choosing $\xi$
orthogonal to the brane, we ensure gauge independency \cite{8} and
have perturbations of the embedding along a single orthogonal
extra direction $\bar{{\cal N}}_{a}$ giving local coordinates of
the perturbed brane as
\begin{eqnarray}\label{a2}
{\cal Z}^{A}_{,\mu}(x^{\nu},\xi^{a}) = {\cal Y}^{A}_{,\mu} +
\xi^{a}\bar{{\cal N}}^{A}_{a,\mu}(x^{\nu}),
\end{eqnarray}
where $\xi^{a}$ $(a = 1,2,...,m-4)$ is a small parameter along
${\cal N}^{A}_{a}$ that parameterizes the extra noncompact
dimensions. One can see from equation (\ref{a1}) that since the
vectors $\bar{{\cal N}}^{A}$ depend only on the local coordinates
$x^{\mu}$, they do not propagate along the extra dimensions
\begin{eqnarray}\label{a3}
{\cal N}^{A}_{a}(x^{\mu}) = \bar{{\cal N}}^{A}_{a} +
\xi^{b}[\bar{{\cal N}}_{b} , \bar{{\cal N}}_{a}]^{A} = \bar{{\cal
N}}^{A}_{a}.
\end{eqnarray}
The above  assumptions lead to the embedding equations of the
perturbed geometry
\begin{eqnarray}\label{a4}
{\cal G}_{\mu \nu }={\cal G}_{AB}{\cal Z}_{\,\,\ ,\mu }^{A}{\cal
Z}_{\,\,\ ,\nu }^{B},\hspace{0.5cm}{\cal G}_{\mu a}={\cal
G}_{AB}{\cal Z}_{\,\,\ ,\mu
}^{A}{\cal N}_{\,\,\ a}^{B},\hspace{0.5cm}{\cal G}_{AB}{\cal N}_{\,\,\ a}^{A}%
{\cal N}_{\,\,\ b}^{B}={g}_{ab}.
\end{eqnarray}
If we set ${\cal N}_{\,\,\ a}^{A}=\delta _{a}^{A}$, the metric of
the bulk space can be written in the following matrix form
\begin{equation}
{\cal G}_{AB}=\left( \!\!\!%
\begin{array}{cc}
g_{\mu \nu }+A_{\mu c}A_{\,\,\nu }^{c} & A_{\mu a} \\
A_{\nu b} & g_{ab}%
\end{array}%
\!\!\!\right) ,  \label{F}
\end{equation}%
where
\begin{equation}
g_{\mu \nu }=\bar{g}_{\mu \nu }-2\xi ^{a}\bar{K}_{\mu \nu a}+\xi ^{a}\xi ^{b}%
\bar{g}^{\alpha \beta }\bar{K}_{\mu \alpha a}\bar{K}_{\nu \beta
b}, \label{G}
\end{equation}%
is the metric of the perturbed brane, so that
\begin{equation}
\bar{K}_{\mu \nu a}=-{\cal G}_{AB}{\cal Y}_{\,\,\,,\mu }^{A}{\cal
N}_{\,\,\ a;\nu }^{B},  \label{H}
\end{equation}%
represents the extrinsic curvature of the original brane (second
fundamental form). We use the notation $A_{\mu c}=\xi ^{d}A_{\mu
cd}$, where
\begin{equation}
A_{\mu cd}={\cal G}_{AB}{\cal N}_{\,\,\ d;\mu }^{A}{\cal N}_{\,\,\ c}^{B}=%
\bar{A}_{\mu cd},  \label{I}
\end{equation}%
represents the twisting vector fields (the normal fundamental form). Any fixed $%
\xi ^{a}$ signifies a new perturbed geometry, enabling us to
define an extrinsic curvature similar to the original one by
\begin{equation}
\widetilde{K}_{\mu \nu a}=-{\cal G}_{AB}{\cal Z}_{\,\,\ ,\mu }^{A}{\cal N}%
_{\,\,\ a;\nu }^{B}=\bar{K}_{\mu \nu a}-\xi ^{b}\left( \bar{K}_{\mu \gamma a}%
\bar{K}_{\,\,\ \nu b}^{\gamma }+A_{\mu ca}A_{\,\,\ b\nu
}^{c}\right) . \label{J}
\end{equation}%
Note that definitions (\ref{F}) and (\ref{J}) require
\begin{equation}
\widetilde{K}_{\mu \nu a}=-\frac{1}{2}\frac{\partial {\cal G}_{\mu \nu }}{%
\partial \xi ^{a}}.  \label{M}
\end{equation}%
In geometric language, the presence of gauge fields $A_{\mu a}$
tilts the embedded family of sub-manifolds with respect to the
normal vector ${\cal N} ^{A}$. According to our construction, the
original brane is orthogonal to the normal vector ${\cal N}^{A}.$
However,  equation (\ref{a4})  shows that this is not true for the
deformed geometry. Let us change the embedding coordinates and set
\begin{equation}
{\cal X}_{,\mu }^{A}={\cal Z}_{,\mu }^{A}-g^{ab}{\cal
N}_{a}^{A}A_{b\mu }. \label{mama40}
\end{equation}%
The coordinates ${\cal X}^{A}$ describe a new family of embedded
manifolds whose members are always orthogonal to ${\cal N}^{A}$.
In this coordinates the embedding equations of the perturbed brane
is similar to the original one, described by equations (\ref{a}),
so that ${\cal Y}^{A}$ is replaced by ${\cal X}^{A}$. This new
embedding of the local coordinates are suitable for obtaining
induced Einstein field equations on the brane. The extrinsic
curvature of a perturbed brane then becomes
\begin{equation}
K_{\mu \nu a}=-{\cal G}_{AB}{\cal X}_{,\mu }^{A}{\cal N}_{a;\nu }^{B}=\bar{K}%
_{\mu \nu a}-\xi ^{b}\bar{K}_{\mu \gamma a}\bar{K}_{\,\,\nu b}^{\gamma }=-%
\frac{1}{2}\frac{\partial g_{\mu \nu }}{\partial \xi ^{a}},
\label{mama42}
\end{equation}%
which is the generalized York relation and shows how the extrinsic
curvature propagates as a result of the propagation of the metric
in the direction of extra dimensions. The components of the
Riemann tensor of the bulk written in the embedding vielbein
$\{{\cal X}^{A}_{, \alpha}, {\cal N}^A_a \}$, lead to the Gauss-
Codazzi equations \cite{12}
\begin{eqnarray}\label{a5}
R_{\alpha \beta \gamma \delta}=2g^{ab}K_{\alpha[ \gamma
a}K_{\delta] \beta b}+{\cal R}_{ABCD}{\cal X} ^{A}_{,\alpha}{\cal
X} ^{B}_{,\beta}{\cal X} ^{C}_{,\gamma} {\cal X}^{D}_{,\delta},
\end{eqnarray}
\begin{equation}\label{ab5}
2K_{\alpha [\gamma c; \delta]}=2g^{ab}A_{[\gamma ac}K_{ \delta]
\alpha b}+{\cal R}_{ABCD}{\cal X} ^{A}_{,\alpha} {\cal N}^{B}_{c}
{\cal X} ^{C}_{,\gamma} {\cal X}^{D}_{,\delta},
\end{equation}
where ${\cal R}_{ABCD}$ and $R_{\alpha\beta\gamma\delta}$ are the
Riemann tensors for the bulk and the perturbed brane respectively.
Contracting the Gauss equation (\ref{a5}) on ${\alpha}$ and
${\gamma}$, we find
\begin{eqnarray}\label{a7}
R_{\mu\nu}=(K_{\mu\alpha c}K_{\nu}^{\,\,\,\,\alpha c}-K_{c} K_{\mu
\nu }^{\,\,\,\ c})+{\cal R}_{AB} {\cal X}^{A}_{,\mu} {\cal
X}^{B}_{,\nu}-g^{ab}{\cal R}_{ABCD}{\cal N}^{A}_{a}{\cal
X}^{B}_{,\mu}{\cal X}^{C}_{,\nu}{\cal N}^{D}_{b},
\end{eqnarray}
which readily gives
\begin{eqnarray}\label{a8}
G_{\mu\nu}=G_{AB} {\cal X}^{A}_{,\mu}{\cal
X}^{B}_{,\nu}+Q_{\mu\nu}+g^{ab}{\cal R}_{AB}{\cal N}^{A}_{a}{\cal
N}^{B}_{b} g_{\mu\nu}- g^{ab}{\cal R}_{ABCD}{\cal N}^{A}_{a}{\cal
X}^{B}_{,\mu}{\cal X}^{C}_{,\nu}{\cal N}^{D}_{b},
\end{eqnarray}
where  $G_{\mu\nu}$ is the Einstein tensor of the brane and
\begin{eqnarray}\label{a9}
Q_{\mu\nu}=-g^{ab}\left(K^\gamma_{\mu a}K_{\gamma\nu b}-K_a
K_{\mu\nu b}\right)+\frac{1}{2}\left(K_{\alpha\beta
a}K^{\alpha\beta a}-K_a K^a\right)g_{\mu\nu}. \label{eqq7}
\end{eqnarray}
As can be seen from the definition of $Q_{\mu\nu}$,  it is
independently a conserved quantity, that is $Q^{\mu\nu}_{;\mu}=0$
\cite{9}. Using the decomposition of the Riemann tensor into the
Weyl curvature, the Ricci tensor and the scalar curvature
\begin{eqnarray}\label{a10}
{\cal R}_{ABCD}=C_{ABCD}-\frac{2}{(m-2)}\left({\cal G}_{B[D}{\cal
R}_{C]A}-{\cal G}_{A[D}{\cal
R}_{C]B}\right)-\frac{2}{(m-1)(m-2)}{\cal R}({\cal G}_{A[D}{\cal
G}_{C]B}),
\end{eqnarray}
we obtain the $4D$ Einstein equations as
\begin{eqnarray}\label{a11}
G_{\mu\nu}&=&G_{AB} {\cal X}^{A}_{,\mu}{\cal
X}^{B}_{,\nu}+Q_{\mu\nu}-{\cal
E}_{\mu\nu}+\frac{m-3}{(m-2)}g^{ab}{\cal R}_{AB}{\cal
N}^{A}_{a}{\cal
N}^{B}_{b}g_{\mu\nu}\nonumber\\
&-&\frac{m-4}{(m-2)}{\cal R}_{AB}{\cal X}^{A}_{,\mu}{\cal
X}^{B}_{,\nu}+\frac{m-4}{(m-1)(m-2)}{\cal
R}g_{\mu\nu},\label{eqq12}
\end{eqnarray}
where
\begin{eqnarray}\label{a12}
{\cal E}_{\mu\nu}=g^{ab} C_{ABCD}{\cal N}^{A}_{a}{\cal
X}^{B}_{,\mu}{\cal N}^D_b{\cal X}^C_{,\nu},
\end{eqnarray}
is the electric part of the Weyl tensor $C_{ABCD}$. Now, let us
write the Einstein equation in the bulk space as
\begin{equation}\label{a13}
G^{(b)}_{AB}+\Lambda^{(b)} {\cal G}_{AB}=\alpha^{*}
S_{AB},\label{eqq14}
\end{equation}
where $\alpha^{*}=\frac{1}{M_{*}^{m-2}}$ ($M_{*}$ is the
fundamental scale of energy in the bulk space), $\Lambda^{(b)}$ is
the cosmological constant of the bulk and $S_{AB}$ consists of two
parts
\begin{equation}\label{a14}
S_{AB}=T_{AB}+ \frac{1}{2} {\cal V}{\cal G}_{AB},\label{eqq15}
\end{equation}
where $T_{AB}$ is the energy-momentum tensor of the matter
confined to the brane through the action of the confining
potential $\cal{V}$. We require $\cal{V}$  to satisfy three
general conditions: firstly, it has a deep minimum on the
non-perturbed brane, secondly, depends only on extra coordinates
and thirdly, the gauge group representing the subgroup of the
isometry group of the bulk space is preserved by it \cite{8}. Use
of equation (\ref{eqq14}) results in
\begin{equation}\label{a15}
{\cal R}=-\frac{2}{m-2}(\alpha^{*} S-m\Lambda^{(b)}),
\end{equation}
and
\begin{equation}\label{a16}
{\cal R}_{AB}=-\frac{\alpha^{*}}{(m-2)}{\cal G}_{AB}
S+\frac{2}{(m-2)}\Lambda^{(b)} {\cal G}_{AB}+\alpha^{*} S_{AB}.
\end{equation}
Substituting ${\cal R}$ and ${\cal R}_{AB}$ from the above into
equation (\ref{a11}), the tangent component of equation
(\ref{eqq14}), also known as the ``gravi-tensor equation'',
becomes
\begin{eqnarray}\label{a17}
G_{\mu\nu}&=& Q_{ \mu\nu} - {\cal E}_{\mu\nu}
+\frac{(m-3)}{(m-2)}\alpha^{*}g^{ab}S_{ab}g_{\mu\nu}+\frac{2\alpha^{*}}{(m-2)}S_{\mu\nu}
- \frac{(m-4)(m-3)}{(m-1)(m-2)}\alpha^{*}Sg_{\mu\nu}\nonumber\\
&+&\frac{(m-7)}{(m-1)}\Lambda^{(b)}g_{\mu\nu}.\label{new1}
\end{eqnarray}
On the other hand, again from (\ref{eqq14}), the trace of the
Codazzi equation (\ref{ab5}) gives the ``gravi-vector equation''
\begin{equation}
K^\delta_{a\gamma;\delta} - K_{a,\gamma} - A_{ba\gamma}K^b +
A_{ba\delta}K^{b\delta} + B_{a\gamma} =
\frac{3(m-4)}{m-2}\alpha^{*}S_{a\gamma},\label{new2}
\end{equation}
where
\begin{equation}
B_{a\gamma} = g^{mn}C_{ABCD}{\cal N}^A_m{\cal N}^B_a{\cal
X}^C_{,\gamma}{\cal N}^D_n.
\end{equation}
Finally, the ``gravi-scalar equation'' is obtained from the
contraction of (\ref{a7}), (\ref{a11}) and using equation
(\ref{a13})
\begin{equation}
\alpha^{*}\left[\frac{m-5}{m-1}S - g^{mn}S_{mn}\right]g_{ab} =
\frac{m-2}{6}\left(Q + R +W\right)g_{ab} -
\frac{4}{m-1}\Lambda^{(b)}g_{ab},\label{new3}
\end{equation}
where
\begin{equation}
W = g^{ab}g^{mn}C_{ABCD}{\cal N}^A_m{\cal N}^B_b{\cal N}^C_n{\cal
N}^D_a.
\end{equation}
Equations (\ref{new1})-(\ref{new3}) represent the projections of
the Einstein field equations on the brane-brane, bulk-brane, and
bulk-bulk directions.

As was mentioned in the introduction, localization of matter on
the brane is realized in this model by the action of a confining
potential. This can simply be realized by
\begin{equation}
\alpha\tau_{\mu\nu} = \frac{2\alpha^{*}}{(m-2)}T_{\mu\nu},
\hspace{.5 cm}T_{\mu a}=0, \hspace{.5 cm}T_{ab}=0,\label{new4}
\end{equation}
where $\alpha$ is the scale of energy on the brane. Now, the
induced Einstein field equation on the original brane can be
written as
\begin{equation}
G_{\mu\nu} = \alpha \tau_{\mu\nu}
-\frac{(m-4)(m-3)}{2(m-1)}\alpha\tau g_{\mu\nu} + \Lambda
g_{\mu\nu} + Q_{\mu\nu} - {\cal E}_{\mu\nu},\label{a8}
\end{equation}
where  $\Lambda= \frac{(m-7)}{(m-1)} \Lambda^{(b)}$. As was noted
before, $Q_{\mu\nu}$ is a conserved quantity which according to
\cite{9} may be considered as an energy-momentum tensor of a dark
energy fluid representing the $x$-matter, the more common phrase
being `X-Cold-Dark Matter' (XCDM). This matter has the most
general form of the equation of state which is characterized by
the following conditions \cite{13}: first it violates the strong
energy condition at the present epoch for $\omega_x<-1/3$ where
$p_x=\omega_x\rho_x$, second, it is locally stable {\it i.e.}
$c^2_s=\delta p_x/\delta\rho_x\ge 0$, and third, causality holds
good, that is $c_s\le 1$. Ultimately, we have three different
types of `matter' on the right hand side of equation (\ref{a8}),
namely, ordinary confined conserved matter represented by
$\tau_{\mu\nu}$, the matter represented by $Q_{\mu\nu}$ which will
be discussed later and finally, the Weyl matter represented by
${\cal E}_{\mu\nu}$.

At this point, it would be appropriate to consider the case where
the bulk space metric can be written as a flat piece plus a
perturbation. This worths looking at since questions like
localization of gravity on the brane and corrections to the
Newtonian potential stems mostly from such a linearized theory. In
the usual brane models the problem of localization of gravity is
discussed in several papers. For example, in \cite{Brevik}, the
authors address the localization of gravity on the
Friedmann-Robertson-Walker (FRW) type branes embedded in either
$AdS_5$ or $dS_5$ bulk space and show that the graviton zero mode
is trapped on the brane. Non-trapped, massive Kaluza-Klein  (KK)
modes correspond to a correction to Newton's law. Also in
\cite{Chaichian}, the Randall-Sundrum model with localized gravity
is considered, replacing the extra compact space-like dimension by
a time-like one. In this way the authors show that the solution to
the hierarchy problem can be reconciled with correct cosmological
expansion of the visible universe. One the other hand in
\cite{Durrer}, the authors show that, in brane models with a
compact extra dimension, like that of the Randall-Sandrum two
brane model, or models with a universal extra dimension
\cite{Feng} which allow for a zero mode like that known from the
KK theories, the five dimensional linearly perturbed Einstein
equations are in conflict with observations.

To briefly comment on linearized gravity in our model, consider
the linear expansion of the bulk metric around the $m$-dimensional
Minkowski spacetime
\begin{equation}
{\cal G}_{AB} = \eta_{AB} + h_{AB}.
\end{equation}
It is straightforward to show that the linear approximation of the
Einstein field equations in the bulk space, in the harmonic gauge,
is
\begin{equation}
\left[ \Box - \frac{4}{m-2}\Lambda^{(b)} - \alpha^{*}{\cal V}
\right]h_{AB} = \frac{4}{m-2}\Lambda^{(b)} \eta_{AB} +
2\alpha^{*}\left( T_{AB} -
\frac{1}{m-2}T\eta_{AB}\right),\label{new5}
\end{equation}
where
\begin{equation}
\Box\equiv \eta^{AB}\partial_A \partial_B = \Box^{(4)} +
\eta^{ab}\partial_a \partial_b.
\end{equation}
One may therefore make a linear expansion of  the metric of the
bulk space and obtain three sets of equations that describe the
gravitons, gravi-vectors and gravi-scalars respectively. The
existence of the confining potential in the linearized field
equations (\ref{new5}) may then have various effects that could be
addressed in a separate work.
\section{Cosmological equations}
In what follows we will analyze the influence of the trace of
$\tau_{\mu\nu}$ and the extrinsic curvature terms on a FRW
universe, regarded as a brane embedded in an $m$ dimensional bulk.
The spatially flat FRW line element is written as
\begin{equation}\label{a18}
ds^2=-dt^2+a(t)^2\left[dr^2+r^2\left(d\theta^2+\sin^2\theta
d\varphi^2\right)\right].
\end{equation}
As the  source we take the perfect fluid given in co-moving
coordinates by
\begin{equation}\label{a19}
\tau_{\mu\nu}=\rho u_{\mu}u_{\nu}+ph_{\mu\nu},\hspace{.5
cm}u_{\mu}=-\delta^{0}_{\mu},\hspace{.5 cm}p=(\gamma-1)\rho,
\end{equation}
where
$$h_{\mu\nu}=g_{\mu\nu}+u_\mu u_\nu.$$
The Weyl tensor ${{\cal E}_{\mu\nu}}$, appearing in equation
(\ref{a8}) is given by
\begin{equation}\label{a20}
{\cal E}_{\mu\nu}=-{\cal U}\frac{\alpha^*}{4}
\left(u_{\mu}u_{\nu}+\frac{1}{3}h_{\mu\nu}+{\cal P}_{\mu\nu}+{\cal
Q}_{\mu}u_{\nu}+{\cal Q}_{\nu}u_{\mu}\right),
\end{equation}
where ${\cal U}$ is an effective nonlocal energy density on the
brane which arises from the gravitational field in the bulk and is
negative  for localizing  the gravitational field near the brane
and reads
\begin{equation}\label{ab21}
{\cal U}=-\frac{4}{\alpha^*}{\cal E}_{\mu\nu}u^{\mu}u^{\nu}.
\end{equation}
Since ${{\cal E}_{\mu\nu}}$ is traceless, its effective local
pressure is $P=\frac{1}{3}{\cal U}$. On the other hand, an
effective nonlocal anisotropic stress is given by
\begin{equation}\label{ab22}
{\cal P}_{\mu\nu}=-\frac{4}{\alpha^*}{\cal E}_{[\mu\nu]},
\end{equation}
while an effective energy flux on the brane is
\begin{equation}\label{ab23}
{\cal Q}_{\mu}=-\frac{4}{\alpha^*}\left({\cal
E}_{\mu\nu}u^{\nu}+{\cal E}_{\nu\mu}u^{\mu}\right).
\end{equation}
The contracted Bianchi identities in the bulk space
$G^{AB}_{\,\,\,\,\,\,\,\,\,;A}=0$, using equation (\ref{eqq14}),
imply
\begin{equation}\label{a21}
\left(T^{AB}+\frac{1}{2} {\cal{V}} {\cal {G}}^{AB}\right)_{ ;A}=0.
\end{equation}
Since the potential $\cal V$ has a minimum on the brane, the above
conservation equation reduces to
\begin{equation}\label{a22}
\tau^{\mu\nu}_{\,\,\,\,\,;\mu}=0,
\end{equation}
and gives
\begin{equation}\label{a23}
\dot{\rho}+3\frac{\dot{a}}{a}\gamma\rho=0.
\end{equation}
This is the conservation equation for the matter fields on the
brane. Taking the covariant derivative of both sides of the
induced Einstein equation (\ref{a8}) and taking into account the
conservation of the matter represented by $Q_{\mu\nu}$ and using
equation (\ref{a22}), we find
\begin{eqnarray}\label{a24}
{\cal E}^{\mu\nu}{;\nu}=\frac{\alpha}{4}g^{\mu\nu}\tau_{,\nu}.
\end{eqnarray}
As we can see from the above equation, $\tau_{\mu\nu}$ is the
source for ${\cal E}_{\mu\nu}$. After substituting from equation
(\ref{a19}) and considering the isotropic form of ${\cal
E}^{\mu\nu}$ from equation (\ref{a20}), we obtain
\begin{equation}\label{a25}
\dot{{\cal U}}+4\frac{\dot{a}}{a}{\cal U}+\rho
\frac{\alpha}{4}\gamma(3\gamma-2)\dot{a} a^{3\gamma+1}=0,
\end{equation}
with solution
\begin{equation}\label{a26} {\cal
U}=\frac{\alpha\gamma\rho}{3a^{3\gamma}}+\frac{c}{a^4}.
\end{equation}
From hereon, we consider an $AdS_m$, $dS_m$ or flat bulk, so that
${\cal E}_{\mu\nu}=0$. For late times this assumption seems
reasonable because the effects of such a term is negligible. The
Codazzi equations (\ref{ab5}) with the assumption of vanishing
twisting vector fields to make the problem at hand simpler, reduce
to
\begin{equation}\label{ab26}
K_{\alpha\gamma a;\sigma}-K_{\alpha\sigma a;\gamma}=0.
\end{equation}
Using the Yorks relation
\begin{equation}\label{a27}
K_{\mu \nu a}=-\frac{1}{2}\frac{\partial
g_{\mu\nu}}{\partial\xi^{a}},
\end{equation}
we realize that in the FRW space-time (diagonal metric),
$K_{\mu\nu a}$ is diagonal. After separating the spatial
components, the Codazzi equations reduce to
\begin{equation}\label{a28}
K_{\mu\nu a,\rho}-K_{\nu\sigma a}\Gamma^{\sigma}_{\mu\rho}=
K_{\mu\rho a,\nu}-K_{\rho\sigma a}\Gamma^{\sigma}_{\mu\nu}.
\end{equation}
\begin{equation}\label{a29}
K_{\mu\nu a,0}-K_{\mu\nu
a}\frac{\dot{a}}{a}=-a\dot{a}\left(\delta^{1}_{\mu}\delta^{1}_{\nu}+r^2
\delta^{2}_{\mu}\delta^{2}_{\nu}+r^2\sin\theta^2\delta^{3}_{\mu}\delta^{3}_{\nu}\right)K_{00
a}.
\end{equation}
The first equation gives $K_{11 a,\nu}=0$  for $\nu\neq1$, since
$K_{11 a}$ does not depend on the spatial coordinates. After
defining $K_{11 a}=b_{a}(t)$, where $b_{a}(t)$ are arbitrary
functions of $t$, the second equation gives
\begin{equation}\label{ab29}
K_{00
a}=-\frac{1}{\dot{a}}\frac{d}{dt}\left(\frac{b_{a}}{a}\right).
\end{equation}
For $\mu,\nu=2,3$ we obtain $K_{22 a}=b_{a}(t)r^2$ and $K_{33
a}=b_{a}(t)r^2\sin^2\theta$ and generally $(\mu,\nu \neq 0)$
\begin{equation}\label{a30}
K_{\mu\nu a}=\frac{b_{a}}{a^2}g_{\mu\nu}.
\end{equation}
We find from (\ref{a9}) that
\begin{equation}\label{a312}
Q_{\mu\nu}=-\frac{1}{a^4}\left(2\frac{b_a
\dot{b}^a}{H}-b_{a}b^{a}\right)g_{\mu\nu} ,\hspace{.5 cm}
Q_{00}=\frac{3 b_{a}b^{a}}{a^4}.
\end{equation}
Assuming that the functions $b_{a}$ are equal and Denoting
$b_{a}=b$, $h=\frac{\dot{b}}{b}$ and $H=\frac{\dot{a}}{a}$, the
components of $Q_{\mu\nu}$ become
\begin{equation}\label{a31}
Q_{\mu\nu}=-\frac{n b^2}{a^4}\left(2\frac{h}{H}-1\right)g_{\mu\nu}
,\hspace{.5 cm} Q_{00}=\frac{3n b^2}{a^4},
\end{equation}
where $n=m-4$. It would how be interesting to see how the above
geometrical interpretation is compared with the X-matter
explanation, a phenomenological candidate for dark energy. To this
end we consider $Q_{\mu\nu}$ as a perfect fluid and write
\begin{equation}\label{a32}
Q_{\mu\nu}=-\frac{1}{\alpha}\left[\rho_{x} u_{\mu}u_{\nu}+p_{x}
h_{\mu\nu}\right],\hspace{.5 cm} p_{x}=(\gamma_{x}-1)\rho_{x}.
\end{equation}
Use of the above equations leads to an equation for $b(t)$
\begin{equation}\label{a35}
\frac{\dot{b}}{b}=\frac{1}{2}\left(4-3\gamma_{x}(t)\right)\frac{\dot{a}}{a}.
\end{equation}
It is interesting to note that this equation resembles one of the
phenomenological candidates for dark energy, the x-matter
\cite{13}, but in our case this field has a fundamental
geometrical justification for the equation of state, having been
derived from the term $Q_{\mu\nu}$ in the Einstein equation
(\ref{a17}), itself a result of the extrinsic curvature. If
$\gamma_{x}$ is taken as a constant, the solution for $b(t)$ is
\begin{equation}\label{a36}
b(t)=b_{0}a(t)^{\frac{1}{2}(4-3\gamma_{x})},
\end{equation}
where $b_{0}$ is an integration constant. With this solution the
energy density of XCDM becomes
\begin{equation}\label{a37}
\rho_{x}=\frac{3n b_{0}^2}{\alpha}a^{-3\gamma_{x}}.
\end{equation}
Using this density for $Q_{\mu\nu}$, the Friedman equations become
\begin{equation}\label{a38}
\frac{\dot{a}^2}{a^2}=\frac{\alpha\gamma}{4}\rho_{0}
a^{-3\gamma}-\frac{\Lambda}{3}+n b_{0}^{2}a^{-3\gamma_{x}},
\end{equation}
\begin{equation}\label{a39}
\frac{\ddot{a}}{a}=-\frac{\alpha\gamma}{4}\rho_{0}a^{-3\gamma}-
\frac{\Lambda}{3}+n b_{0}^2
a^{-3\gamma_{x}}\left(1-\frac{3}{2}\gamma_{x}\right).
\end{equation}
A qualitative classification of the solutions on the basis of
different values of the parameter $\gamma_{x}$ can be achieved
without solving these equations. If one defines the potential
\begin{equation}\label{a40}
V(a)=-\frac{\alpha\gamma}{4}\rho_{0} a^{-3\gamma+2}-n
b_{0}^{2}a^{-3\gamma_{x}+2},
\end{equation}
then equation (\ref{a38}) may be written as
\begin{equation}\label{a41}
\dot{a}^2+V(a)=0.
\end{equation}
The qualitative behavior of the scale factor $a(t)$ for different
values of $\gamma_{x}$ may be realized from the above equation by
noting that $\dot{a}^2$ is positive. This behavior is much
dependent on the range of the values that $\gamma_{x}$ can take.
We distinguish the following possibility for having an
accelerating universe
\begin{equation}\label{a46}
0 <\gamma_{x}<\frac{2}{3}.
\end{equation}

The behavior of the potential $V(a)$ and the corresponding
evolution of the scale factor $a(t)$ are illustrated in figure 1.
\begin{figure}
\centerline{\begin{tabular}{ccc}
\epsfig{figure=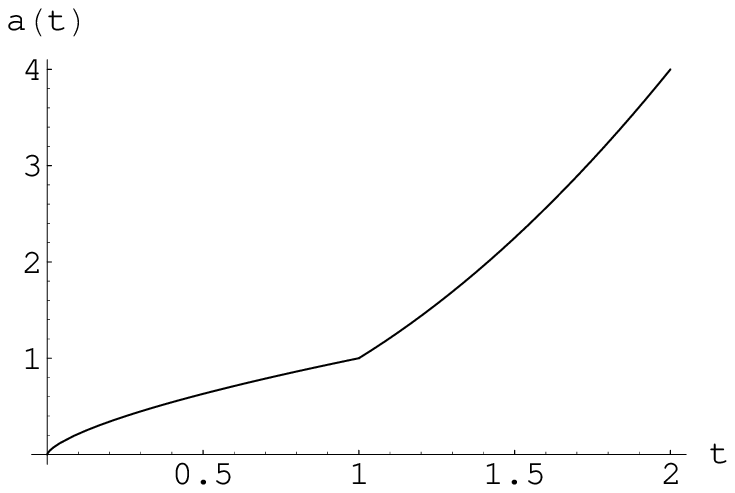,width=6cm}\hspace{2cm}
\epsfig{figure=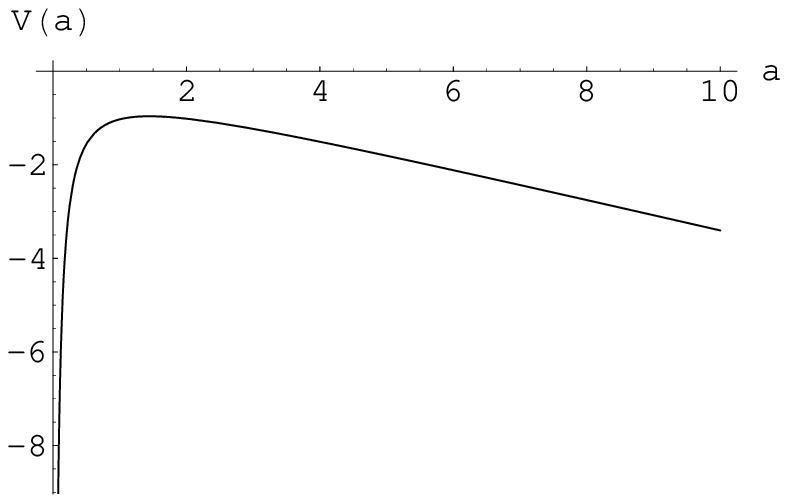,width=6cm}
\end{tabular} }
\caption{\footnotesize Left, the behavior of the scale factor
$a(t)$ and right, the potential $V(a)$ for $\gamma=1$ and
$\gamma_{x}=\frac{1}{3}$.} \label{fig1}
\end{figure}
\section{The Accelerating universe}
It is well known that in FRW cosmology accelerated expansion of
the universe may only be obtained in the case when the universe is
filled with some exotic form of matter giving rise to a negative
pressure, {\it e.g.} a cosmological constant. However, if we look
at figure \ref{fig1} we note that, within the context of the
present model, the universe can also exhibit accelerated expansion
in the case of a vanishing or positive pressure. Unfortunately,
the Friedmann equation $\dot{a}^{2}+n
b_{0}^{2}a=\frac{\alpha\varrho_{0}}{4a}$ (derived from (\ref{a38})
for $\gamma=1$ and $\gamma_{x}=\frac{1}{3}$ ) cannot be solved in
closed form. However, in two extreme cases corresponding to small
and large $a(t)$, exact solutions may easily be found
\begin{equation}\label{a49}
a(t)=\left(\frac{9\alpha\gamma\rho_{0}}{16}\right)^{\frac{1}{3}}t^\frac{2}{3},
\hspace{.5 cm} \mbox{for small} \hspace{2mm}a,
\end{equation}
and
\begin{equation}\label{a50}
a(t)=\frac{n b_{0}^{2}}{4}t^{2}, \hspace{.5 cm}  \mbox{for large}
\hspace{2mm}a.
\end{equation}
The first  solution is of the Einstein de-Sitter type, while the
second represents an evidently inflationary of the power-law type.
This means that in our model the universe starts as decelerating
and finally ends up as accelerating. In the simplest FRW
cosmological models with a one-component fluid filling up the
universe such behavior is not possible.  The declaration parameter
for the model reads
\begin{equation}\label{a51}
q=-\frac{\ddot{a}a}{\dot{a}^2}=\left(3\gamma_{x}-2\right)\frac{\Omega_{x}}{2}
+\frac{3}{4}\Omega_{m},
\end{equation}
where $\Omega_{m}=\frac{\alpha\rho_{m}}{3H^2}$ and
$\Omega_{x}=\frac{\alpha\rho_{x}}{3H^2}$. For $\Omega_{m}\sim 0.3$
and $\Omega_{x}\sim0.7 $, as has been suggested by recent
observations, the present epoch requires $\gamma_{x}<0.52$ as in
the x-matter scenarios \cite{13}. In the next section we discuss
the observational parameters of the model.
\section{The Age of universe}
Long existing discrepancy between a relativity  large value of the
Hubble parameter  $H\sim701\mbox{cm}/ \mbox{sec}/\mbox{mpc}$  and
the large universe age is nicely resolved if it existed
$\Omega_{x}\sim0.7$. If we compare two regimes of cosmological
expansion, acceleration and deceleration, then with the same value
of the hubble parameter at the present epoch, expansion was slower
in the past for the accelerating regime. It means that to reach
the same magnitude of $H_{0}$ more time was necessary and the
accelerating universe should be older. We find the age of the
universe  by direct integration of the Freidmann equation
(\ref{a38}),
\begin{equation}\label{a52}
t_{0}^{B}=\frac{1}{H_{0}}\int^{1}_{0}\frac{dx}{\left(\frac{\Omega_{m}}{x}+
\Omega_{x}x\right)^\frac{1}{2}},
\end{equation}
Analogously, the calculated age of the universe in FRW models
reads
\begin{equation}\label{a523}
t_{0}^{F}=\frac{1}{H_{0}}\int^{1}_{0}\frac{dx}{\left[\frac{\Omega_{m}}{x}+
(1-\Omega_{0})\right]^\frac{1}{2}},
\end{equation}
where $H_{0}^{-1}=9.8 \times 10^9 \mbox{h}^{-1}$ years and the
dimensionless parameter $\Omega_{x}$, according to modern data, is
about 0.7. Hence, in the flat matter dominated universe with
$\Omega_{0}=1$ the age of the universe would be only 9.3 Gyr,
whereas the old globular clusters indicate much larger age 12-15
Gyr. On the other hand in our model for flat universe with
$\Omega_{m}=0.3$ and $\Omega_{x}=0.7$ the age of the universe,
according to equation (\ref{a52}), is 12 Gyr,  in good agreement
with the range quoted above. We have plotted the age of the
universe in both models as a function of the energy density
parameter in figure 2.
\begin{figure}
\centerline{\begin{tabular}{c} \epsfig{figure=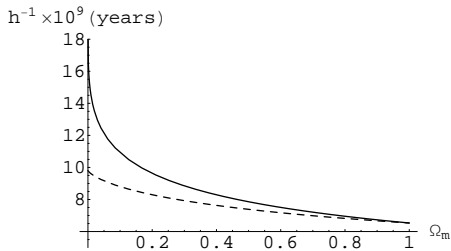,width=6cm}
\end{tabular} }
\caption{\footnotesize Age of the universe as predicted by the
present model, solid line, and the usual FLRW cosmological model,
dashed line.} \label{fig1}
\end{figure}
\section{Conclusions}
In this paper we have studied a brane world model in which the
matter is confined to the brane through the action of a confining
potential, rendering the use of any junction condition redundant.
This has provided the ground for presenting a scenario in which a
FRW universe is embedded in an $m$ dimensional bulk where the
extrinsic curvature causes the universe to accelerate. This result
could be of interest since we have shown that the existence of
exotic matter (given as having a negative pressure) is not
necessary to drive an accelerated expansion. Finally, we have
found that the age of the universe in this model is remarkably
larger than the FRW models.


\begin{thebibliography}{99}
\bibitem{1} T. Shiromizu, K. Maeda, and M. Sasaki,
{\it Phy. Rev.} D {\bf 62} (2000) 024012  (gr-qc/9910076).
\bibitem{2} L. Randall and R. Sundrum,   {\it Phys. Rev. Lett.}
{\bf 83} (1999) 4690 (hep-th/9906064).
\bibitem{3}  P. Brax and C. van de Bruck,
{\it Class. Quantum. Grav.} {\bf 20} (2003) R201-R232
(hep-th/0303095)
\bibitem{4} R. Maartens, Reference Frames and Gravitomagnetism, ed. J Pascual-Sanchez et al. (World Sci., 2001), p93-119
(gr-qc/0101059).
\bibitem{5} D. Langlois,   {\it Prog. Theor. Phys. Suppl.} {\bf
148} (2003) 181 (hep-th/0209261).
\bibitem{6} R. A. Battye and B. Carter,  {\it Phys. Lett.} B {\bf 509}
(2001) 331  (hep-th/0101061).
\bibitem{7} V. A. Rubakov and M. E. Shaposhnikov, {\it Phys. Lett.} B {\bf 125},
(1983) 136.
\bibitem{8} S. Jalazadeh and H. R. Sepangi, {\it Class. Quant. Grav.}
{\bf 22} (2005) 2035 (gr-qc/0408004).
\bibitem{9} M. D. Maia, E. M. Monte, J. M. F. Maia and J. S.
Alcaniz, {\it Class. Quantum Grav.} {\bf 22} (2005) 1623
(astro-ph/0403072).
\bibitem{10} M. D. Maia, Edmundo M. Monte, J. M. F. Maia, {\it Phys.
Lett.} B {\bf 585} (2004) 11 (astro-ph/0208223).
\bibitem{12}L. P. Eisenhart  1966 {\it Riemannian Geometry}, (Princeton University
Press).
\bibitem{13} M. Turner and M. White, {\it Phys. Rev. } D {\bf 56}
(1997) 4439,\\T. Chiba, N. Sugiyama and T. Nakamura, {\it Mon.
Not. R. Astron. Soc.} {\bf 289} (1997) L5.
\bibitem{Brevik}I. Brevik, K. Ghoroku, S. D. Odintsov and M.
Yahiro, Phys. Rev. D {\bf 66} (2002) 064016.
\bibitem{Chaichian}M. Chaichian and A. B. Kobakhidze, Phys. Lett.
B {\bf 488} (2000) 117.
\bibitem{Durrer}R. Durrer and P. Kocian, Class. Quant. Grav. {\bf
21} (2004) 2127.
\bibitem{Feng} J. L. Feng, A. Rajaraman and F. takayama, Phys.
Rev. D {\bf 68} (2003) 08518.
\end{thebibliography}
\end{document}